# TWO COMPONENTS OF DEPOLARIZATION CURRENTS IN PVDF CAUSED BY RELAXATION OF HOMO- AND HETEROCHARGE

S. N. Fedosov, A. F. Butenko, and A. E. Sergeeva
Odessa National Academy of Food Technologies, Odessa, Ukraine

*The procedure has been developed for extracting homocharge and heterocharge currents from experimentally measured thermally stimulated depolarization currents of corona poled PVDF. Application of different depolarization modes supplemented with the isothermal currents allowed to obtain such parameters of the relaxation processes, as activation energies, characteristic frequencies and time constants.*

### Introduction

Polyvinylidene fluoride (PVDF) and its co-polymers received considerable attention during last years because of their high piezo- and pyroelectric activity whose origin is not fully understood [1–8]. Their specific properties are usually attributed to the high level of the residual polarization [4–7], although some researchers believe that the injected charge can also play an important role [8].

Despite the fact that PVDF is often considered as a polymer ferroelectric [5], some of its electrical properties can be explained in terms of the conventional theory of polar dielectrics and electrets. The phenomenological Gross-Swann-Gubkin model of electrets [9–11] assumes availability of two kinds of charges in polar dielectrics, namely the homocharge $s(t)$ whose sign coincides with the polarity of electrodes during poling and the heterocharge $?(t)$ (internal polarization) arising from micro- and macrodisplacement of intrinsic charges in the dielectric under action of the electric field. In the case of PVDF one can assume that the heterocharge represents the dipole polarization, while the homocharge is formed by charges trapped at or near the surface [12].

Stability of the electret state in a polar dielectric depends on the interaction and the resulting mutual relaxation of the homocharge and the heterocharge. Since the heterocharge (polarization) is usually of the primary importance for PVDF, the role of the homocharge was not given enough consideration so far, although the stabilizing effect of the space charge on the residual polarization has been already discussed [12, 13].

Thermally stimulated depolarization (TSD) is a commonly used method for identifying relaxation processes in charged polymer electrets [14–18]. However, it is very difficult to separate the influence of the homocharge and the heterocharge on the TSD currents, especially if the corresponding peaks are superimposed in a wide range of temperatures.

In the present work, we illustrate how to extract depolarization currents caused by relaxation of the homocharge and the heterocharge from the total TSD current by solving the inverse problem and revealing the relaxation behavior of both components from the experimentally measured TSD current. Moreover, it is shown that application of different TSD modes supplemented with the isothermal depolarization currents makes it possible to evaluate important parameters of the relaxation processes, such as activation energies, characteristic frequencies and time constants.

### Experimental procedure

The study was performed on uniaxially stretched 25 μm thick PVDF films supplied by Plastpolymer (Russia) and composed of *ß*-form crystallites and amorphous phase in nearly equal volume fractions. A metal electrode of 0.1 μm thickness was deposited on one surface of each sample by thermal evaporation of Al in a vacuum. The other side of the sample was subjected to a negative corona discharge initiated by a pointed tungsten electrode with the automatically controlled potential, while the metallized rear surface was grounded. The vibrating control grid between the sample surface and the corona electrode was kept at a constant potential of 3 kV. All samples were charged at room temperature under a constant charging current density [19–20] of 90 µA/m² for 30 min and then short-circuited and conditioned at room temperature for 24 hours (except those intended for measuring kinetics of the electret potential).

Four versions (modes) of depolarization were applied to study relaxation processes, namely thermally stimulated (T) and isothermal (I) depolarization of short-circuited (S) and open-circuited (O) samples. The modes were thus referred as TS, TO, IS and IO modes with the first letter indicating the temperature regime (thermally stimulated or isothermal) and the second one indicating the electric state (short-circuited or open-circuited). Additional experiments on the thermally stimulated kinetics of the electret potential (TP) have been also performed after 24 hours of conditioning in the open circuit configuration. The Teflon film of 10 µm thickness was used as a dielectric gap in TO and IO modes. All thermally stimulated experiments were performed under a constant heating rate of 3 K/min. In isothermal experiments, temperature was maintained constant after the desired value was achieved by fast heating. The electret potential in the TP mode was measured by the Kelvin method and recorded continuously.

### Results and their discussion

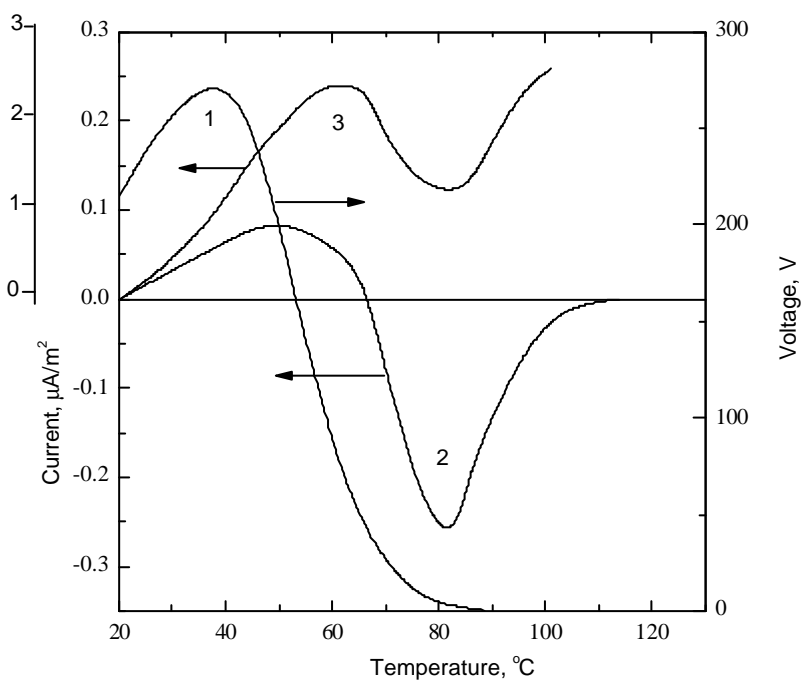

Fig. 1 Thermally stimulated currents in the TS mode (1) and the TO mode (2) and the electret potential in the TP mode (3).

The following are the main features of experimental curves seen in Fig. 1 and Fig. 2
– The depolarization current in the TS mode shows a broad "non-classical" peak with a maximum around 65°C;
– The inversion of the TSD current is observed in the TO mode, while the direction of the current coincides with that in the TS mode during the initial stage of the heating;
– The electret potential in the TP mode has a maximum at 40 °C.
– The current slowly decreases with time in the IS mode at all temperatures, while the isothermal current changes its direction in the IO mode at elevated temperatures.

The above mentioned features can be explained in the frames of a model assuming existence in the samples of the homocharge and the heterocharge [1, 9–11] with the former representing the charge trapped at the surface and the latter standing for the polarization formed in the bulk due to the electric field created by the homocharge. The two types of the charge are obviously interdependent.

At first we examine charging and relaxation processes qualitatively. It is reasonable to assume that the negatively charged particles (ions and/or electrons), supplied by corona discharge are adsorbed and thermalized on the surface of the sample because of their low (thermal) energy. The redundant charge localized on the surface or in the near-to-surface layer forms the homocharge having a certain surface density $s$ and producing the uniform field $E$ in the bulk of the sample. The high electron affinity of the fluorine atoms facilitates the charge trapping and formation of the stable homocharge.

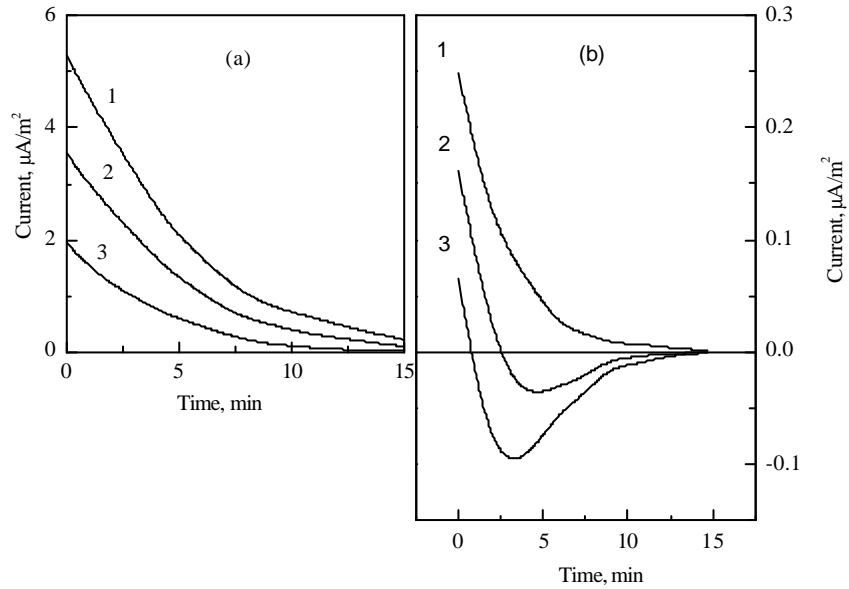

Fig. 2 Isothermal transient currents at different temperatures in the IS mode (a) and the IO mode (b);
1 - 45 °C, 2 - 55 °C, 3 - 70 °C.

The uniform internal polarization $P$ (heterocharge) appears mainly due to alignment of $CH_2$-$CF_2$ dipoles in the field created by the homocharge. This is equivalent to developing the bound surface charge $P$ having the sign opposite to that of the homocharge $s$. Among all polarization processes in PVDF, the alignment of $CH_2$-$CF_2$ dipoles is the main one, because of their large dipole moment of 2.1 D [1, 2].

If the polarization $P$ is zero, the field in the bulk of the sample is created by the total surface charge $s$. When polarization $P$ starts to grow, the depolarizing field appears, which is immediately "neutralized" by a fraction of the surface charge equal to the neutralized (screened) polarization. Thus, the electric field in the bulk is now created not by the total charge $s$, but by the difference ($s-P$) between the surface charge and the polarization. Hence, one can consider the total surface charge $s$ as consisting of two parts $s=s_1+s_2$, the first one representing the compensating charge ($s_1=P$) and the second one $s_2=s-P$ creating the electric field in the bulk of the sample.

After short-circuiting of the charged samples (in the TS and the IS modes), the "excessive" charge $s_2$ disappears. Then the equilibrium between the homocharge and the heterocharge ($s=s_1=P$), as well as zero internal field ($E=0$) are maintained due to the current flowing through the external circuit, so the measured current corresponds to the relaxation of the heterocharge.

However, if, after the short-circuiting and forming the $s=s_1=P$ equilibrium, a non-conductive dielectric gap is introduced between one of the electrodes and the surface of the sample (TO and IO modes), one can observe the relaxation currents of both, the heterocharge and the homocharge flowing in opposite directions. The field in the bulk is not zero any more, so the surface charge (homocharge) is either forced to drift in its own field from one surface of the sample to another one through the whole thickness of the sample (in the absence of the intrinsic conductivity), or it is slowly neutralized by charge carriers responsible for the intrinsic conductivity. In any case, the relaxation of the heterocharge takes place in non-zero field conditions and is caused by thermal disordering of the aligned dipoles [1, 14].

It will be shown now that the two components of the total depolarization current can be obtained from the experimental $i(T)$ dependence in the TO mode (Fig. 1, curve 2). It is known [1, 14] that the TSD current $i(t)$ and the electret potential $V(t)$ in experiments with a non-conducting spacer or an air gap introduced between the sample surface and the electrode depend not only on the interrelation between the homocharge and the heterocharge, but also on their time derivatives, so that

$$i(t) = s\left[\frac{dP(t)}{dt} - \frac{d\mathbf{s}(t)}{dt}\right], \tag{1}$$

$$V(t) = \frac{sx_1}{\mathbf{e}_o\mathbf{e}_1}[\mathbf{s}(t) - P(t)], \tag{2}$$

$$i(t) = -\frac{\mathbf{e}_o\mathbf{e}_1}{x_1} \cdot \frac{dV(t)}{dt}, \tag{3}$$

where $s = x_o\mathbf{e}_1/(x_1\mathbf{e} + x_o\mathbf{e}_1)$, $t$ is the time, $\mathbf{e}$ and $x_o$ the dielectric constant and the thickness of the sample, $\mathbf{e}_1$ and $x_1$ the corresponding values of the dielectric gap, $\mathbf{e}_o$ is the permittivity of a vacuum. The conductive component $i_c(t)$ of the total current can be expressed as

$$i_C(t) = \frac{g}{x_o}V(t) = -\frac{d\mathbf{s}(t)}{dt}, \tag{4}$$

where $g = g_o\exp(-Q/kT)$ is the specific conductivity, $k$ Boltzmann's constant, $T$ the temperature, $Q$ the activation energy of the intrinsic conductivity, $g_o$ the pre-exponential factor. Integrating Eq.(3) and substituting time $t$ for temperature $T$ in Eq. (1)–(4) according to $T = T_o(1+bt)$, where $b$ is the heating rate, $T_o$ the initial temperature, we obtain expressions for temperature dependences of the homocurrent $i_1(T)$ and the heterocurrent $i_2(T)$, as well as for the voltage across the sample (electret potential) $V(T)$

$$i_1(T) = \frac{d\mathbf{s}}{dt} = -\frac{x_1 g_o}{bT_o x_o \mathbf{e}_o \mathbf{e}_1}\exp\left(-\frac{Q}{kT}\right)\int_T^\infty i(T')dT', \tag{5}$$

$$i_2(T) = \frac{dP}{dt} = \frac{i(T)}{s} + \frac{d\mathbf{s}}{dt}, \tag{6}$$

$$V(T) = \frac{x_1}{bT_o\mathbf{e}_o\mathbf{e}_1}\int_T^\infty i(T')dT'. \tag{7}$$

All the values at the right hand side of Eq. (5)–(7) are known, or can be found experimentally. Results of the calculations according to the Eq.(5)–(7) based on the data of Fig. 1 are shown in Fig. 3. Values of $Q=0{,}76$ eV and the pre-exponential factor $g_o=0{,}18$ $O^1 \cdot m^{-1}$ were obtained from the steady-state values of the isothermal charging currents and voltages.

As one can see from Fig. 3, the homocurrent and the heterocurrent form two broad peaks with almost coinciding maxima. The heterocharge decays faster in the low-temperature region, while the homocharge remains relatively stable. This is probably the reason of the initial increase of the thermally stimulated potential (see curve 3 in Fig. 1 and curve 3 in Fig. 3). The current inversion in TO and IO modes is caused by the change of ratio between the homocurrent and the heterocurrent at high temperatures (curves 1 and 2 in Fig. 3).

It is known that the inversion of the TSD current can be caused by the over-polarization, i.e. by appearing of the additional heterocharge in the field of the homocharge, the voltage in this case should be decreasing [14]. However, this has not been observed experimentally in our case (Fig. 1). From the other side, the initial growth of the electret potential during the heating cannot be caused by increasing of the surface charge density $s$, because the charges in this case should move in the direction opposite to that of the electric field created by the charges, that is not possible. Therefore, the first TSD peak and the corresponding increase of the electret potential (Fig. 1) are caused by the faster decay of the heterocharge (polarization) in comparison with the homocharge. It is possible that in PVDF not all polarization is destroyed during the first stage of the heating, but only the least stable part.

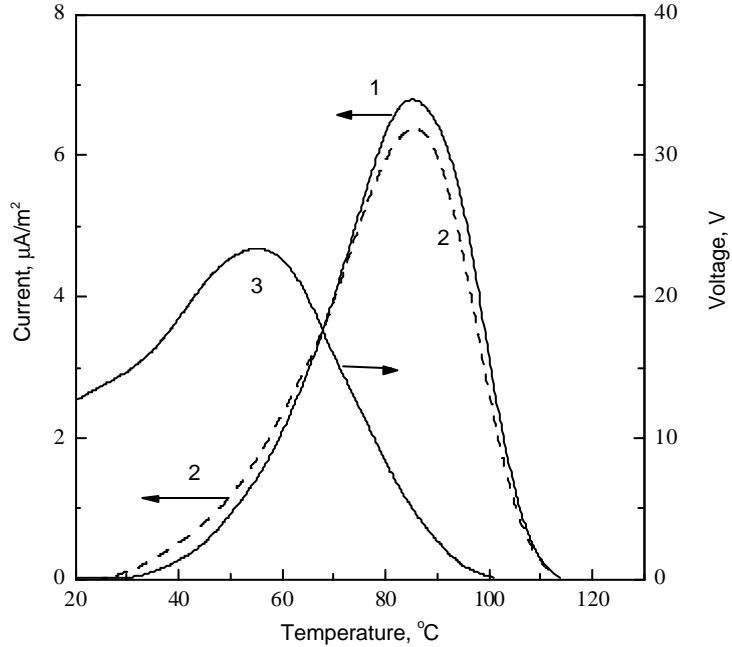

Fig. 3 Temperature dependences of the homocurrent (1), the heterocurrent (2) and the voltage across the sample (3) calculated according to the model.

Thus, the heterocharge in PVDF films is not sufficiently stable, therefore its lasting conservation is possible only in presence of the stabilizing field of the homocharge. We believe that many specific properties of PVDF are related to a fortunate combination of the large dipole moment of $CH_2$–$CF_2$ units (2.1 D) [2] promoting formation of the important heterocharge and the high electron affinity of the fluorine atoms (3.37 eV) promoting creation of the stable homocharge. Although the electret state in PVDF is unstable, the self-balanced mutual relaxation of the homocharge and the heterocharge is delayed due to the stabilizing action of the homocharge.

It is assumed in the theory of electrets [1,9-11] that the homocharge and the heterocharge decay exponentially with the temperature dependent time constants. Therefore, the following expressions are valid for IO and IS modes

$$i_1(t) = -\frac{s\mathbf{s}_o}{\mathbf{t}_1}\exp\left(-\frac{t}{\mathbf{t}_1}\right), \tag{8}$$

$$i_2(t) = -\frac{P_o}{\mathbf{t}_2}\exp\left(-\frac{t}{\mathbf{t}_2}\right), \tag{9}$$

$$\mathbf{t}_1(T) = \frac{\mathbf{e}_o \mathbf{e}}{g_o}\exp\left(\frac{Q}{kT}\right), \tag{10}$$

$$\mathbf{t}_2(T) = \mathbf{t}_o \exp\left(\frac{W}{kT}\right), \tag{11}$$

where $W$ is the activation energy of the heterocharge relaxation, $t_1$ and $t_2$ the corresponding time constants.

Applying Eq.(8)–(11) to the experimental curve shown in Fig. 2, we calculated the following parameters of the homocharge and heterocharge relaxations: the activation energies ($Q$=0,76 eV and $W$=0,54 eV), the characteristic frequencies ($f_2 = 1/t_o$ =7,4 MHz and $f_1 = (g_o/e_o e)$=1,7 GHz), the time constants at 20°C ($t_1$ =31 000 s and $t_2$ =2 800 s). The results indicate that the homocharge is much more stable than the heterocharge.

**Conclusion**

It is shown how to extract depolarization homocharge and heterocharge currents from experimentally measured TSD current and reveal the relaxation behavior of the both components. The application of different TSD modes supplemented with the isothermal depolarization currents allowed to find important parameters of the relaxation processes, such as activation energies, characteristic frequencies and time constants.

The uniform field approximation assumed in this paper is justified only in the case of high poling fields exceeding 50–60 MV/m. At lower fields, one should consider injection of charge carriers in the bulk resulting in non-uniformity of the field and the polarization in the thickness direction [6, 7].

The proposed method allows to analyze interrelation between the homocharge and the heterocharge not only in PVDF, but also in other dielectrics. Introduction of polar groups with simultaneous creation of deep traps for the charge carriers might be a promising procedure for increasing stability of the residual polarization in polar polymer dielectrics. Therefore, if the appropriate conditions exist for creating and trapping of the homocharge, then the high level of the residual polarization can also be ensured for a long period of time.